\documentstyle[twoside,fleqn,espcrc2,amsbsy,latexsym,epsf]{article}

\newcommand{\bra}[1]{\langle #1 \vert}
\newcommand{\ket}[1]{\vert #1 \rangle}
\newcommand{\mod}[1]{\vert {#1}\vert}

\title{The Glue Around Quarks and the Interquark Potential}

\author{E. Bagan\address{Grup de F\'{\i}sica Te\`{o}rica, Departament
                                                      de F\'{\i}sica
and
                         IFAE,\\
                         Edifici Cn, Universitat Aut\`{o}noma de
Barcelona,\\
                         E-08193 Bellaterra (Barcelona),
                         Spain}$\!\!\!$,$\,\,$
      R. Horan$^{\rm b}$,  M. Lavelle$^{\rm b}$\thanks{Talk presented by M.\ Lavelle.}
        and
        D. McMullan\address{School of Mathematics and
                            Statistics,\\
                            University of Plymouth, Plymouth,\\
                            PL4 8AA, United Kingdom}
                          }

\begin{document}

\begin{abstract}
The quarks of quark models cannot be identified with the quarks of
the QCD Lagrangian. We review the restrictions that gauge field
theories place on any description of physical (colour) charges.
A method to construct charged particles is presented. The
solutions are applied to a variety of applications. Their Green's
functions are shown to be free of infra-red divergences to all
orders in perturbation theory. The interquark potential is
analysed and it is shown that the interaction responsible
for anti-screening results from the force between two separately
gauge invariant constituent quarks. A fundamental limit on the
applicability of quark models is identified.
\end{abstract}

\maketitle

\section{LONG RANGE FORCES}

The discovery of asymptotic freedom showed that  non-abelian
gauge theory could explain the observed short distance
scaling behaviour. Thus QCD was able to describe partonic physics.
However, it has still not been able to explain how the original
constituent quarks of Gell-Mann emerge from the underlying
dynamics of our theory of the strong interactions. In this talk we
will review how charges should be described in quantum field
theory and see what lessons can be learned from QCD for quark
models.

In the standard textbook description of scattering it is assumed
that at asymptotic times the coupling in some sense vanishes. This
is, however, known not to hold for theories like QED and QCD
where, as a consequence of the masslessness of the gauge bosons,
there is a long range interactions. For Electrodynamics it has
been shown \cite{kulish:1970} that at large times the interaction has the
form $ -{\mathrm{e}}\int\!
d^3x\, A^\mu(x)J^{\mathrm{as}}_\mu(x)$, where the
asymptotic current is
\begin{equation}
J^{\mathrm{as}}_\mu(x)=\int\!
d^3p\,\rho(p)\frac{p^\mu}{E_p}\delta^{(3)}(\underline{x}-\frac{t
\underline{p}}{E_p})\,.
\end{equation}
This shows that even asymptotically we cannot in general \lq turn
the interaction off\rq. The neglect of these interactions leads to
infra-red divergences. Note though that giving the photon a small
mass does cause the interaction Hamiltonian to vanish at large
times --- this further
indicates the intimate connection between the long
range nature of the interaction, gauge invariance and the infra-red problem.

What we would like to stress is that the non-vanishing of the
asymptotic interaction means that Gauss' law is not trivial at
large times. Gauss' law is the generator of local gauge
transformations, so this then implies that the charged matter fields of
the QED and QCD Lagrangians do not become physical in this limit,
i.e., the matter fields are \textit{not} good asymptotic fields.
They must be surrounded, dressed, by an electromagnetic cloud
(gluonic for colour charges).

Being physical in a gauge theory  requires, at the very least,
being locally gauge invariant. We thus write a physical charge
generically as: $h^{-1}\psi$, where $h^{-1}$ is a field
dependent element of the gauge group and we demand that under a
gauge transformation, where the matter transforms as
$\psi\to e^{i{\mathrm{e}\theta}}\psi$, the \textit{dressing}
around the charged matter transforms as $h^{-1}\to e^{
-i{\mathrm{e}\theta}}h^{-1}$.

This requirement is, however, not sufficient to specify the form
of the cloud around a quark: many such dressings can be
constructed (for a review see \cite{india,Lavelle:1997ty}). We need a further
condition to single out physically relevant descriptions. For an
asymptotic particle with four velocity $u^\mu$, we demand that
$h^{-1}$ satisfies the following dressing equation: $
h u^\mu\partial_\mu(h^{-1})=-i{\mathrm{e}}u^\mu A_\mu$
This equation may be motivated \cite{india} both by seeing that in the
heavy quark effective theory it yields a particle with a sharp
momentum and also by the realisation that a  particle dressed in
this manner obeys a free asymptotic dynamics.

In QED it is possible to solve these two demands (gauge invariance
and kinematical) and we then find that for an asymptotic charge
the correct form of the dressing is
\begin{equation}
 h^{-1}=e^{-i{\mathrm{e}}K} e^{-i{\mathrm{e}}\chi}
\end{equation}
where
\begin{eqnarray}
&&K(x)=-\int_{\Gamma}ds(\eta+v)^\mu\frac{\partial^\nu
F_{\nu\mu}}{{\mathcal G}\cdot\partial}\\
&&\chi(x)=\frac{{\mathcal G}\cdot A}{{\mathcal G}\cdot\partial}
\end{eqnarray}
and ${\mathcal G}^\mu=(\eta+v)^\mu(\eta-v)\cdot\partial-\partial^\mu$
with $\Gamma$ the past (future) trajectory of an incoming (outgoing)
particle. Here we have written the four velocity $u^\mu$ as
$\gamma(\eta+v)^\mu$ with $\eta$ the unit temporal vector and
$v=(0,\underline{v})$.

This structured dressing has a gauge dependent part, $\chi$,
which makes the whole charge locally gauge invariant, and is thus
in some sense a minimal dressing. The additional term, $K$,
is gauge invariant but is necessary if we are to fulfil the
dressing equation. To understand this solution better we have
studied the infra-red  behaviour of the Green's functions of
these dressed charges.

In a series of detailed perturbative
tests \cite{Bagan:1997su,Bagan:1998kg} we have
verified to all orders in perturbation theory that these Green's
functions are free of on-shell infra-red divergences. The role of
the two structures in the dressing is noteworthy: the minimal
dressing removes the soft divergences while the additional term,
$K$, kills off phase divergences. These tests are compelling
evidence for the validity of our underlying requirements and for
our ensuing descriptions of charges.

We also stress here that the use of an incorrect dressing does not
yield infra-red finite on-shell Green's functions --- gauge
invariance is not enough: one must find the physically relevant
solutions and their correct interpretation.

\section{COLOUR CHARGES AND CONSTITUENTS}

In QCD physical quarks must be surrounded by a gluonic cloud. This
requirement of gauge invariance is, however, still more urgent in
Chromodynamics. This is because  the colour charge
operator, although not itself gauge invariant, is invariant on locally
gauge invariant states \cite{Lavelle:1996tz}. Thus \textit{any} description
of coloured quarks \textit{must} be gauge invariant. Our other
kinematical requirement we still impose so as to have quarks with
a sharp momentum.

We have constructed  dressings for quarks and gluons
in perturbation theory \cite{Lavelle:1994xa,Lavelle:1997ty}.
The construction of the dressing
appropriate to a static quark may be performed, using a
simple algorithm, to any order in perturbation theory
\cite{Lavelle:1997ty}. However, the minimal requirement of gauge invariance
means that there is a close link between dressings and gauge
fixings. One can thus prove \cite{Lavelle:1997ty} that there is a
topological obstruction to the construction of gauge invariant
quarks with a well defined colour. Quarks are not true QCD
observables, but rather only possess a limited domain of validity.
This domain is the realm of quark models and its breakdown, the
onset of confinement, needs to be precisely
pinned down. This will require non-perturbative calculations
of the interplay between dressings and the non-trivial topology
of QCD. The remainder of this talk though will be dedicated to
a perturbative study \cite{Lavelle:1998dv} ascertaining
whether the quark language
can actually be motivated from QCD.

\section{HADRONS MADE FROM QUARKS?}

It is not just that experimentalists see hadrons rather than
quarks --- some models of hadrons do not involve constituent quarks
at all. We therefore now want to study whether the construction of
colour singlet hadrons from (dressed, gauge invariant) quarks is
energetically favoured or not.  In QED it can be shown that a flux
tube picture of an $e^+e^-$ system is unstable and that it decays
into two separate dressed charges\footnote{See
\texttt{http://www.ifae.es/\~{}roy/qed.html} for an animation of this}.
What we will now do is to construct two dressed static quarks,
separated by some distance $r$, study the potential energy of
this state and compare it with the known QCD result for the
potential of the lowest energy state involving two static quark
(matter) fields.

This potential is known from Wilson loop calculations to have
at order $\alpha^2$ the form:
\begin{equation}
V(r)= -\frac{g^2 C_F}{4\pi r}\left[ 1+
\frac{g^2}{4\pi}\frac{C_A}{2\pi }\left( {4}-{\frac13} \right)
\log(\mu r) \right]\,,
\end{equation}
i.e., a Coulombic potential with an anti-screening logarithmic
correction. This last we have divided up into two parts: an
anti-screening term (the 4) and a screening (the 1/3)
contribution.

The coefficient of the logarithm is proportional to the one-loop
beta function. It is well known that the all-important
anti-screening contribution comes from longitudinal glue, while
the term which tries to screen colour charge arises from gauge
invariant glue \cite{Hughes:1980ms,Nielsen:1978rm}. Our dressings,
as we have seen, are built
out of a minimal part, $\chi$, constructed out of longitudinal
degrees of freedom and the additional gauge invariant term. We
would therefore expect that if in this perturbative region a
description of hadrons in terms of constituent quarks is
energetically favoured, then the minimal dressing should yield the
dominant anti-screening part of the potential.

At lowest order the Coulombic potential is essentially abelian, so
to test the non-abelian nature of hadrons we need to work
at $O(\alpha^2)$. We then require the minimal dressing to
$O(g^3)$. The potential will then be the separation dependent part
of the expectation value of the Hamiltonian between our
constituent quarks.

The abelian minimal dressing for a static charge, $\exp(-ie\chi)$, with
$\chi={\partial_i A_i}/{\nabla^2}$, may be extended to QCD as follows
\begin{equation}
\exp(-ie\chi)\Rightarrow \exp(g\chi^aT^a)\equiv h^{-1}
\end{equation}
with $g\chi^aT^a=(g\chi_1^a+g^2 \chi_2^a+g^3\chi_3^a+\cdots)T^a$
where
\begin{eqnarray}
\chi_1^a&=&\frac{\partial_j A_j^a}{\nabla^2}\,;\\
\chi_2^a&=&f^{abc}\frac{\partial_j}{\nabla^2}\left(
\chi_1^bA_j^c+\frac12(\partial_j\chi_1^b)\chi^c_1
\right)\,;\quad\!\!\dots
\end{eqnarray}
Henceforth $h^{-1}$ signifies this minimal dressing.

The expectation value of the Yang-Mills Hamiltonian, between
such minimally dressed quark/antiquark states,
$\bar\psi(y)h(y)h^{-1}(y')\psi(y')\ket0$ gives the
potential
 \begin{eqnarray}\label{potty}
  -\textrm{tr}\int\!d^3x \,\bra0 &&
 \!\!\!\!\! \!\!\!\!\!
 [E^a_i(x),h^{-1}(y)]h(y)\nonumber\\
&&\; \times [E^a_i(x),h^{-1}(y')]h(y') \ket0\,.
\end{eqnarray}
The equal-time commutator
$[E_i^a(x),A_j^b(y)]=i\delta_{ij}\delta^{ab}
\delta(\boldsymbol{x}-\boldsymbol{y})$ then yields the expected
Coulomb potential:
\begin{equation}
  V^{g^2}(r)=
  -\frac{g^2 N C_F}{4\pi r}\,.
\end{equation}

To test the non-abelian nature of the dressed quarks, we first calculate
$[E^a_i(x),h^{-1}(y)]h(y)$. This may rapidly be  shown to reduce to
\begin{eqnarray}
 g[E_i^a(x),\chi_1(y)]&\!\!\!
 \!\!\!\!\!\!\!&\!\!\!\!\!\!\!\!\!\!\!\!+ g^2[E_i^a(x),\chi_2(y)]\nonumber\\ &&
 \!\!\!\!\!\!\!\!\!\!+g^3 [E_i^a(x),\chi_3(y)]
 +O(g^4)\,,
\end{eqnarray}
\noindent
plus many gauge dependent terms which
cancel amongst themselves in physical quantities. Substituting this
into (\ref{potty}) yields for the potential \cite{Lavelle:1998dv}
\begin{eqnarray}
 V^{g^4}(r)&\!\!=\!\!&
\! - \frac{3g^4 C_FC_A}{(4\pi)^3}\int\!d^3z\!\int\!d^3w
 \frac1{\mod{\boldsymbol z-\boldsymbol w}} \\
&& \!\!\!\!\!\!\!\!\!\!\!\!\!\!\!\!\!\!\!\!\!\! \times
\left(  \partial_j^z\frac1{\mod{\boldsymbol z-\boldsymbol y}}\right)
 \left(  \partial_k^w\frac1{\mod{\boldsymbol w-\boldsymbol {y}'}}\right)
\langle A_k^T(w)A_j^T(z)\rangle\,. \nonumber
\end{eqnarray}
At this order in $\alpha$ we may use the free propagator
\begin{equation}
\langle A_k^T(w)A_j^T(z)\rangle=
\frac1{2\pi^2}\frac{(z-w)_j
(z-w)_k}{{\mod{\boldsymbol z-\boldsymbol w}}^4}\,.
\end{equation}
The integrals are straightforward and we obtain
\begin{equation}
V^{g^4}(r)=-\frac{g^4}{(4\pi)^2}\frac{NC_FC_A}{2\pi
r}\,{4}\log(\mu r)\,,
\end{equation}
which is just the predicted anti-screening contribution to the
potential. This demonstrates that the anti-screening effect
characteristic of non-abelian gauge theories results from the
interaction of two constituent quarks.

\section{CONCLUSIONS}

Gauge theories are characterised by long ranged interactions and
this results in the asymptotic fields obeying a non-trivial
dynamics. This, with the help of Gauss' law, tells us in turn that
physical charged particles, be they electrons or quarks, cannot be
naively identified with the matter fields of the Lagrangian of
our gauge theory. Rather we must use dressed fields, where the
cloud surrounding any physical charge is taken into account.
In this talk we have presented a scheme to construct dressed
charges with well-defined velocities.

In explicit QED calculations we saw that these dressings solve the
infra-red problem already at the level of Green's functions. The
dressings have two components which separately cancel the two
sorts of infra-red singularities in QED. We note that in the theory of
massless electrodynamics there is evidence \cite{india} that the
collinear divergences typical of theories with massless charges
will also be removed by the solutions to the dressing equation.

The need for dressings is most blatant in QCD. We simply cannot
talk about coloured particles in any meaningful way without them.
If coloured quarks obeying Pauli statistics are to emerge from
the the underlying Lagrangian as a description of hadrons
they will be dressed quarks.

Perturbatively we have a well-defined method for constructing
dressings that yield gauge invariant quarks, but, as we saw above,
there is a global obstruction (essentially the Gribov ambiguity)
preventing the construction of an observable quark. This sets a
\textit{fundamental limit on the applicability of quark models}.

However, even when quarks can still be constructed, they might
not be physically significant. It could be energetically preferred
to construct colour singlet, gauge invariant hadrons in a fashion
that does not involve constituent quarks. We have therefore
calculated the potential felt by two gauge invariant quarks
in a colour singlet state using minimally dressed quarks. This led
directly to the the anti-screening contribution found in the
QCD lowest energy state with two matter fields. This shows that
this paradigm non-abelian effect is due to the interaction of two
gauge invariant quarks. Furthermore the calculation presented here
precisely identifies the gluonic configuration responsible for
this force.

The immediate next steps are to identify the glue underlying the
screening part of the potential (is this the additional $K$
term in the dressing?) and to study the anti-screening
contribution at higher orders in perturbation theory.
Beyond perturbation theory, we recall that the global limitation
on constructing gauge invariant quarks means that any constituent
picture of hadrons will break down once sufficient
non-perturbative physics is probed. The fundamental aim of the
programme of research sketched in this talk is to determine the
scale at which the apparent constituent structure of hadrons
is revealed to be a mirage.

\medskip
\noindent{\bf Acknowledgments:}
This work was partly supported by the British Council and the Acciones
Integradas  programme (HB 1997-0141).

%\bibliographystyle{h-physrev}
%\bibliography{litbank1}

\end{document}